\documentclass[3p]{elsarticle}
\makeatletter
\def\ps@pprintTitle{%
 \let\@oddhead\@empty
 \let\@evenhead\@empty
 \def\@oddfoot{\centerline{\thepage}}%
 \let\@evenfoot\@oddfoot}
\makeatother

\usepackage{amssymb}

\usepackage{graphicx}
\usepackage{hyperref}
\hypersetup{
    breaklinks=true,   
    colorlinks=true,   
    pdfusetitle=true,  
} 

\usepackage{amsmath}
\usepackage{placeins}
\usepackage{subfigure}
\usepackage{tikz}
\usepackage[normalem]{ulem} 
\usepackage[makeroom]{cancel} 
\usetikzlibrary{shapes.geometric, arrows}
\usetikzlibrary{positioning}

\newcommand{\X}{\mathbf{X}}

\begin{document}

\begin{frontmatter}

\title{Minimum Time Control of a Gantry Crane System with Rate Constraints}

\author[mysecondaryaddress]{Adrian Stein}

\author[mysecondaryaddress]{Tarunraj Singh}
\cortext[mycorrespondingauthor]{Corresponding author}
\ead{tsingh@buffalo.edu}

\address[mysecondaryaddress]{Department of Mechanical and Aerospace Engineering, University at Buffalo (SUNY),\\ Buffalo, NY 14260-4400, USA}

\begin{abstract}
This paper focuses on the development of minimum time control profiles for point-to-point motion of a gantry crane system in the presence of uncertainties in modal parameters. Assuming that the velocity of the trolley of the crane can be commanded and is subject to limits, an optimal control problem is posed to determine the bang-off-bang control profile to transition the system from a point of rest to the terminal states with no residual vibrations. Both undamped and underdamped systems are considered and the variation of the structure of the optimal control profiles as a function of the final displacement is studied. As the magnitude of the rigid body displacement is increased, the collapse and birthing of switches in the optimal control profile are observed and explained. Robustness to uncertainties in modal parameters is accounted for by forcing the state sensitivities at the terminal time to zero. The observation that the time-optimal control profile merges with the robust time-optimal control is noted for specific terminal displacements and the migration of zeros of the time-delay filter parameterizing the optimal control profile are used to explain this counter intuitive result. A two degree of freedom gantry crane system is used to experimentally validate the observations of the numerical studies and the tradeoff of increase in maneuver time to the reduction of residual vibrations is experimentally illustrated.
\end{abstract}


\end{frontmatter}

\section{Introduction}\label{sec:intro}
Control of cranes is a topic that has garnered increased interest over the past three decades coinciding with the growth in the use of prefiltering approaches to minimize residual vibrations of systems characterized by underdamped motion. 
A vast majority of crane controllers can be classified as open-loop or closed-loop, with a few combining feedforward and feedback controllers in a tracking framework.
One open-loop approach is called input shaping \cite{Singer.1990} which consists of a time-delay filter which is designed to cancel the underdamped poles of the system \cite{Singh.2010}. The domain of input shaping has matured and can account for uncertainties in model parameters. To account for uncertainties in the estimated damping or natural frequencies of the underdamped poles, multiple zeros of the time-delay filter are placed at the nominal locations of the underdamped poles, resulting in robustness to uncertainties in the modal parameters. Controllers which are robust around the nominal model~\cite{Singer.1990} and those that account for interval domains of uncertainties~\cite{Singh.2002} have been developed. Constraints on jerk~\cite{Singh.2004} and deflection~\cite{Singhose.1997} have also been taken into account in the design. Recently distributed delay input shapers~\cite{Vyhlidal.2013} have been studied which introduce a novel parameterization in the design of input shapers.
Including input shapers within a feedback loop has also been considered~\cite{Staehlin.2003, Pilbauer.2015} as researchers explore techniques to exploit the strengths of input shapers. 

Noakes, Petterson, and Werner~\cite{Noakes.1990} proposed a switching control profile to generate
oscillation-damped transport and swing-free stop. Their technique consists of bang-off-bang
acceleration profiles in which the pulses are timed to minimize the cable sway during the
maneuver and results in a swing-free stop. They experimentally demonstrated the results of
the open-loop control design. Shah and Hong~\cite{Shah.2014} applied input shaping for the underwater transport of nuclear power plant’s fuel. There have been numerous publications related to the use of input shapers~\cite{Singer.1990, Singh.2010, Hong.2019} for sway control of cranes~\cite{Garrido.2008, Singhose.1997, Singhose.2008, Singhose.2000, Matsui.2022, FasihurRehman.2022,Cuong.2021, Jaafar.2021, Zhang.2021, Zhang.2020, Zhang.2022, Sun.2014}.
Maghsoudi et al.~\cite{Maghsoudi.2017} applied a distributed time-delay filter on a gantry crane. They used the method proposed by Vyhlidal et al.~\cite{Vyhlidal.2013} for uncertainty studies for gantry crane control and demonstrated that applying a distributed time-delay filter lead to an asymmetric robustness behaviour of the residual sway about the nominal stage. Ramli et al.~\cite{Ramli.2018} designed a neural network-based input shaper while Yavuz and Beller~\cite{Yavuz.2021} used neural networks for a closed-loop controller when it is combined with an input shaper. Wahrburg et al.~\cite{Wahrburg.2022} and Ramli et al.~\cite{Ramli.2020} applied input shaping for maneuvers of an overhead crane with non-zero initial conditions, with an objective of zero residual oscillations of the payload. Additionally, work has been done in command shaping control for non-zero initial and final conditions \cite{Alhazza.2022}. Stein and Singh~\cite{Stein.2022} presented simulation results of velocity constrained design of input shaped control profiles for a gantry crane system. Fliess et al.~\cite{Fliess.1991} used the concept of differential flatness to control the traversing and hoisting of an overhead crane. This differential flatness based design was extended to discrete time design by Diwold et al.~\cite{Diwold.2022}.

 Alli and Singh~\cite{Alli.1999} designed passive controllers for a distributed parameter representation of the crane cable for point-to-point maneuvers where the integral of the time absolute error is minimized. 
O'Connor~\cite{OConnor.2003} used a wave equation representation of the cable dynamics assuming that the velocity of the trolley could be commanded. The velocity of the trolley was assumed to be constrained and the damping was assumed to be zero. Researchers included particle swarm optimization into controller design for overhead cranes~\cite{Azmi.2019, Maghsoudi.2019}. Golovin et al.~\cite{Golovin.2019} developed a $H_{\infty}$ robust controller for actively damping the structural vibrations of the gantry crane system.
Various control methods have also been proposed where the overhead crane is modelled as a double pendulum system~\cite{Jaafar.2019, Huang.2015, Wu.2020, Mar.2017}.
A few papers proposed controllers which combine closed-loop controller used in conjunction with open-loop shaped profiles to track. Kolar et al.~\cite{Kolar.2017} proposed a hybrid solution combining an open-loop generated crane trajectory as a reference signal and closed-loop controller for handling external disturbances. 
 Li et al.~\cite{Li.2019} introduced an online planning method for minimum-time control of overhead cranes. Furthermore, many papers considered the payload as a point mass whereas Stein and Singh~\cite{Stein.2022b} proposed an input shaper used in conjunction with a proportional-derivative controller for a crane with an inertial payload. Other work has considered sliding mode~\cite{Wu.2021, Zhang.2021, Zhang.2019}, adaptive control~\cite{Chen.2019}, discrepancy-based control~\cite{Golovin.2022} and compared different control strategies~\cite{MirandaColorado.2019} while including various external disturbances on cranes. Apart from time-optimal control, Sun et al.~\cite{Sun.2018} investigated an energy-optimal controller for an underactuated double pendulum crane with state and control constraints. Compared to active vibration suppression of a crane's payload, Yurchenko et al.~\cite{Yurchenko.2021} used a passive method with an absorber.

This paper considers a tabletop gantry crane system driven by stepper motors which permits commanding the position of the trolley. By imposing velocity limits on the trolley motion, this paper considers the design of velocity constrained time-optimal point-to-point control of a crane moving in two dimensions. Since the pendular motion is almost undamped, an undamped system model was first considered. Subsequently, the structure of switching function was used to parameterize the bang-off-bang profile and the resulting nonlinear programming problem was solved to determine the optimal solution for any arbitrary maneuver. The bang-off-bang control structure was then generalized to cater to multi-mode systems with underdamped modes. 

The main contributions of this work include: (1) Development of a velocity constrained time-optimal control profile for a gantry crane which is robust to uncertainties in modal parameters, (2) Illustration of the non-intuitive result that the robust and non-robust solutions are coincident for specific displacements, (3) Illustration that a rectangular pulse input can attenuate the dominant vibratory modes for specific displacements, (4) Illustration of the change in structure of the optimal control profile, and (5) Experimental validation of all the aforementioned observations.

Section~\ref{sec:prob_st} presents the development of velocity constrained time-optimal control for an undamped gantry crane system followed by the development of controllers which are robust to uncertainties in the undamped frequencies in Section~\ref{sec:robcon}. Section~\ref{sec:robdesign} generalizes the optimal control formulation for a multi-mode system with damped or undamped modes. Section~\ref{sec:sw_tran} presents a simple approach to determine the transition in the structure of the control profile, followed by the validation of the design on a two degree of freedom gantry crane system in Section~\ref{sec:experiment}. The paper concludes with a brief summary of the results of the paper.
\section{Undamped System}\label{sec:prob_st}
\begin{figure}[thpb]
\centering
	\includegraphics[width=0.35\textwidth]{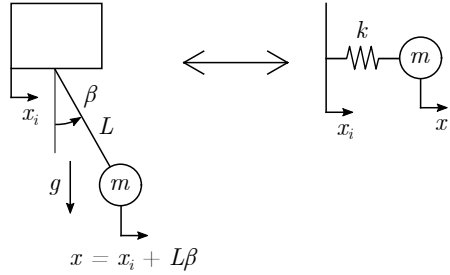}
	\caption{Equivalent spring-mass system for small angle displacements.}
 \label{fig:01}
\end{figure}
\begin{figure}[thpb]
		\centering
		\includegraphics[width=0.48\textwidth]{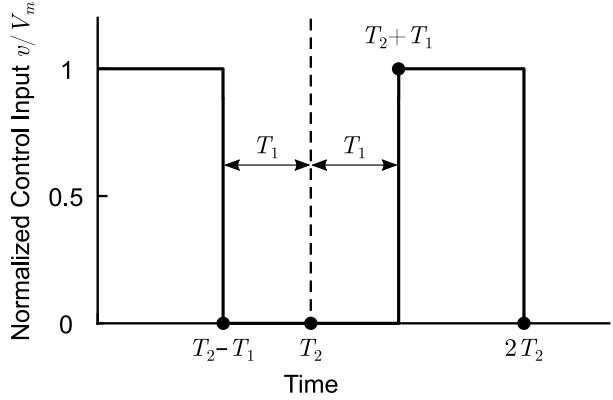}
		\caption{Anti-symmetric time-optimal control profile.}
		\label{fig:02}
\end{figure}

The gantry crane setup includes a trolley driven by a stepper motor which permits the command of the trolley's velocity by assuming that the acceleration is zero as the commanded velocity transitions. A schematic of the crane and an equivalent spring-mass system are shown in Fig.~\ref{fig:01} where a small angle displacement is assumed~\cite{Zhang.2014}~\cite{Chen.2016b}. The spring-mass model can be written as:
\begin{align}
    mL\ddot{\beta}(t) + mg\sin\left(\beta(t)\right) &= 0\\
    m\ddot{x}(t) - m\cancelto{0}{\ddot{x}_i(t)} + mg\left(\frac{x(t)-x_i(t)}{L}\right) &= 0\\
    \leftrightarrow \color{black} m\ddot{x}(t) + kx(t) - kx_i(t) &= 0\\
    \dot{x}_i(t) &= v(t)
\end{align}
where $v$ the velocity of the trolley is considered as the input and is constrained $0 \le v\le V_m$. The assumption that velocity of the trolley can be used as a control input is based on papers which demonstrate velocity input control of industrial cranes~\cite{Sorensen.2007, Suksabai.2020, Knierim.2010, Wu.2021}. Mass normalization leads to the state space equation:
\begin{equation}
\underset{\dot{\X}}{\underbrace{\begin{bmatrix}\dot{x}_1 \\ \dot{x}_{2} \\ \dot{x}_3 \end{bmatrix}}} = 
	\underset{A}{\underbrace{\begin{bmatrix} 0 & 1 & 0 \\ -\omega_n^2 & 0 & \omega_n^2 \\ 0 & 0 & 0 \end{bmatrix}}} \underset{\X}{\underbrace{\begin{bmatrix} x_1 \\ x_2 \\ x_3 \end{bmatrix}}} + 
	\underset{B}{\underbrace{\begin{bmatrix} 0 \\ 0 \\ 1 \end{bmatrix}}} v
\end{equation}
where $\omega_n = \sqrt{\frac{k}{m}}$. The time-optimal control problem can be posed as:
\begin{subequations}
	\begin{align}
	& \mbox{min} ~~J = \int_{0}^{t_f} dt \\
	\mbox{subject to} & \notag \\ & \dot{\X} = A\X+Bv \\
	& \X(0) = \begin{Bmatrix} 0 & 0 & 0 \end{Bmatrix}^T  \\
	&  \X(t_f) = \begin{Bmatrix} x_f & 0 & x_f \end{Bmatrix}^T \\
	& 0 \le v \le V_m \: \: \: \: \forall t
	\end{align}
\end{subequations}
Defining the Hamiltonian\index{Hamiltonian} as:
\begin{equation}
\mathcal{H} = 1+\lambda^{T} \left( A\X+Bv \right),
\end{equation}
the necessary conditions for optimality can be derived using calculus of variations,
resulting in the equations
\begin{subequations} \label{eq:topt_4}
	\begin{align}
	& \dot{\X} = \frac{\partial \mathcal{H}}{\partial \lambda} = A\X+Bv \\
	& \dot{\bf{\lambda}} = -\frac{\partial \mathcal{H}}{\partial \X} = -A^{T} \bf{\lambda} \label{eq:costates}\\
	& v = V_m \mathbf{H}(-\it{B}^T\bf{\lambda}) \label{eq:toptc}\\
	& \X(0) = 0 \mbox{ and } \X(t_f) = \begin{Bmatrix} x_f & 0 & x_f \end{Bmatrix}^T \\
	& \mathcal{H} = 0 \mbox{  at   } t=0
	\end{align}
\end{subequations}
where $\mathbf{H}$ is the Heaviside step function. Eq.~\eqref{eq:toptc} is derived using {\it Pontryagin's minimum principle (PMP)} which requires the optimal trolley velocity to be bang-off-bang. Eq.~\eqref{eq:costates} can be solved in closed form resulting in the equation:
\begin{equation}
\bf{\lambda}(t) = e^{-\it{A}^T t}\bf{\lambda}(0)
\end{equation}
which can also be written as:
\begin{align}
\lambda_1(t) &= \cos(\omega_n t) \lambda_1(0) + \omega_n \sin(\omega_n t) \lambda_2(0) \\
\lambda_2(t) &= -\frac{\sin(\omega_n t)}{\omega_n} \lambda_1(0) + \cos(\omega_n t) \lambda_2(0) \\ 
\lambda_3(t) &= (1-\cos(\omega_n t)) \lambda_1(0) - \omega_n \sin(\omega_n t) \lambda_2(0) + \lambda_3(0) \label{eq:lam3sol}
\end{align}
for undamped systems.
Since the switching function is $\it{B}^T\bf{\lambda}$, it reduces to the third costate which reveals that the switching function is a non-zero mean harmonic. This structure will help comprehend the change in structure of the optimal control profile $v(t)$ as a function of the final displacement $x_f$. Since the Hamiltonian $\mathcal{H}$ at all times, including the initial time, is zero, we have:
\begin{align}
\mathcal{H}(0) & = 1+\lambda^{T} \left( A\X(0)+BV_m \right) = 0\\
\Rightarrow \lambda_3(0) & = -\frac{1}{V_m}. 
\end{align}
Further, for a rest-to-rest maneuver, which is the focus of this paper, the Hamiltonian $\mathcal{H}$ at the final time is:
\begin{align}
\mathcal{H}(t_f) & = 1+\lambda^{T} \left( A\X(t_f)+BV_m \right) = 0\\
\Rightarrow \lambda_3(t_f) & = -\frac{1}{V_m}. \label{eq:lam3tf}
\end{align}
Substituting Eq.~\eqref{eq:lam3tf} into Eq.~\eqref{eq:lam3sol}, we can show that:
\begin{equation}
\frac{\lambda_2(0)}{\lambda_1(0)} = \frac{1}{\omega_n}\tan\left( \frac{\omega_n t_f}{2}\right). \label{eq:lam_con}
\end{equation}
Evaluating the slope of $\lambda_3(t)$ at the initial time and final time, we can show that:
\begin{equation}
\dot{\lambda}_3(0) = -\omega_n^2 \lambda_2(0), \mbox{  and } \dot{\lambda}_3(t_f) = \omega_n^2 \lambda_2(0),
\end{equation}
which permits us to conclude that the switching function is anti-symmetric about the mid-maneuver time, which implies there will always be an even number of switches and that pairs of switch times are equally distant from the mid-maneuver time. Using Eq.~\eqref{eq:lam_con}, the time derivative of $\lambda_3(t)$ evaluated at the mid-maneuver time is:
\begin{align}
\dot{\lambda}_3\left(\frac{t_f}{2}\right) &= \omega_n \sin\left(\omega_n \frac{t_f}{2}\right) \lambda_1(0) - \omega_n^2 \cos\left(\omega_n \frac{t_f}{2}\right) \lambda_2(0) = 0
\end{align}
which implies that the slope of the switching curve is always zero at the mid-maneuver time.
This prompts parameterising the optimal control profile as:
\begin{align}
v(t) &= V_m \left( 1 - \mathbf{H}(t-(T_2-T_1)) +  \mathbf{H}(t-(T_2+T_1)) - \mathbf{H}(t-2T_2)\right)
\end{align}
as illustrated in Fig.~\ref{fig:02},
which in the frequency domain is:
\begin{equation}
V(s) = \frac{V_m}{s} \underset{G_c(s)}{\underbrace{\left( 1 - e^{-s(T_2-T_1)} +  e^{-s(T_2+T_1)} -  e^{-2sT_2}\right)}}
\end{equation}
where $G_c(s)$ is the transfer function of the time-delay filter which generates the bang-off-bang control profile when subjected to a step input. By requiring that a pair of zeros of $G_c(s)$ cancel the undamped poles of the system and the pole at the origin, one can derive the constraint to formulate a parameter optimization problem. It can be seen that:
\begin{equation}
G_c(s=0) = 1-1+1-1=0
\end{equation}
which implies that the transfer function has a zero at the origin. The rigid body boundary condition is determined by integrating the bang-off-bang velocity profile leading to the equation:
\begin{align}
x_i(t) &= V_m\left( t - (t-T_2+T_1)\mathbf{H}(t-T_2+T_1) + (t-T_2-T_1)\mathbf{H}(t-T_2-T_1) -  (t-2T_2)\mathbf{H}(t-2T_2) \right) 
\end{align}
which at the final time of $t=t_f=2T_2$ leads to the equation:
\begin{align}
x_i(2T_2) = & x_f =  V_m\left( 2T_2 - (T_2+T_1) +  (T_2-T_1) \right)\\
\leftrightarrow x_f = & 2V_m (T_2-T_1) \Rightarrow (T_2-T_1) = \frac{x_f}{2V_m}.
\end{align}
Furthermore, to cancel the undamped poles at $s=\pm j\omega_n$, we have:
\begin{align}
G_c(s=j\omega_n) &= 1 - e^{-j\omega_n (T_2-T_1)} +  e^{-j\omega_n(T_2+T_1)} -  e^{-2j\omega_n T_2}=0
\end{align}
which reduces to:
\begin{align}
&1-\cos(\omega_n (T_2-T_1))+\cos(\omega_n(T_2+T_1)) - \cos(2\omega_n T_2) =0 \\
&\sin(\omega_n (T_2-T_1))-\sin(\omega_n(T_2+T_1)) + \sin(2\omega_n T_2) =0
\end{align}
which simplifies to:
\begin{align}
2\sin(\omega_n T_2)\left(\sin(\omega_n T_2) - \sin(\omega_n T_1) \right) &=0 \\
2\cos(\omega_n T_2)\left(\sin(\omega_n T_2) - \sin(\omega_n T_1) \right) &=0
\end{align}
which leads to the solution:
\begin{equation}
T_2 = \frac{\pi}{\omega_n} -T_1
\end{equation}
which results in the closed form solution:
\begin{equation}
2T_2 = \frac{\pi}{\omega_n} + \frac{x_f}{2 V_m} \mbox{  and  } T_1 = \frac{\pi}{2\omega_n} - \frac{x_f}{4 V_m}
\end{equation}
where $2T_2$ is the maneuver time $t_f$. Since both $T_1$ and $T_2$ are functions of the maneuver $x_f$, the scenario of collapse of the switches requires $T_1=0$ which results in the constraint:
\begin{equation}
x_f = \frac{2\pi V_m}{\omega_n}
\label{eq:first_degen}
\end{equation}
which is a bang profile or a constant velocity profile. It should be noted that the collapse of the switches is proportional to one period of the switching function $\frac{2\pi}{\omega_n}$. Any increase in time associated with a maneuver greater than the bang profile will introduce two additional switches. This results from the fact that there will be two peaks or troughs of the switching profile, which results in four switches. An observation which will be exploited later is the fact that since the switching function is a harmonic with a bias, the width of all the off zones in the bang-off-bang profiles will be the same. Consider a constant velocity profile with no zero zones which can be parameterized as:
\begin{equation}
v(t) = V_m \left( 1 - \mathbf{H}(t-2T_2)\right)
\end{equation}
which can be represented in the frequency domain as:
\begin{equation}
V(s) = \frac{V_m}{s} \underset{G_c(s)}{\underbrace{\left( 1 -  e^{-2sT_2}\right)}}.
\end{equation}
The final displacement is given by the equation:
\begin{equation}
x_f = 2V_m T_2.
\label{eq:findisp}
\end{equation}
The requirement that the transfer function $G_c(s)$ places a zero at the location of the undamped poles of the system $s=\pm j\omega_n$ requires the constraints:
\begin{align}
1-\cos(2\omega_n T_2) & = 0 \\
-\sin(2\omega_n T_2) & = 0
\end{align}
which results in the solution:
\begin{equation}
T_2 = \frac{n\pi}{\omega_n}, \mbox{  where } n=1,2,3,\hdots.
\end{equation}
Substituting the solution $T_2$ into Eq.~\eqref{eq:findisp}, we have:
\begin{equation}
x_f = \frac{2n\pi V_m}{\omega_n}.
\end{equation}
The number of zero zones $n$ depends on the desired terminal displacement and is given by the constraint:
\begin{equation}
\frac{2(n-1)\pi V_m}{\omega_n} \le x_f \le \frac{2n\pi V_m}{\omega_n}.
\end{equation}
For example, when $0 \le x_f \le \frac{2\pi V_m}{\omega_n}$, one zero velocity zone exists in the optimal velocity profile.
\begin{figure*}[thpb]
	\centering
	\includegraphics[width=0.96\textwidth]{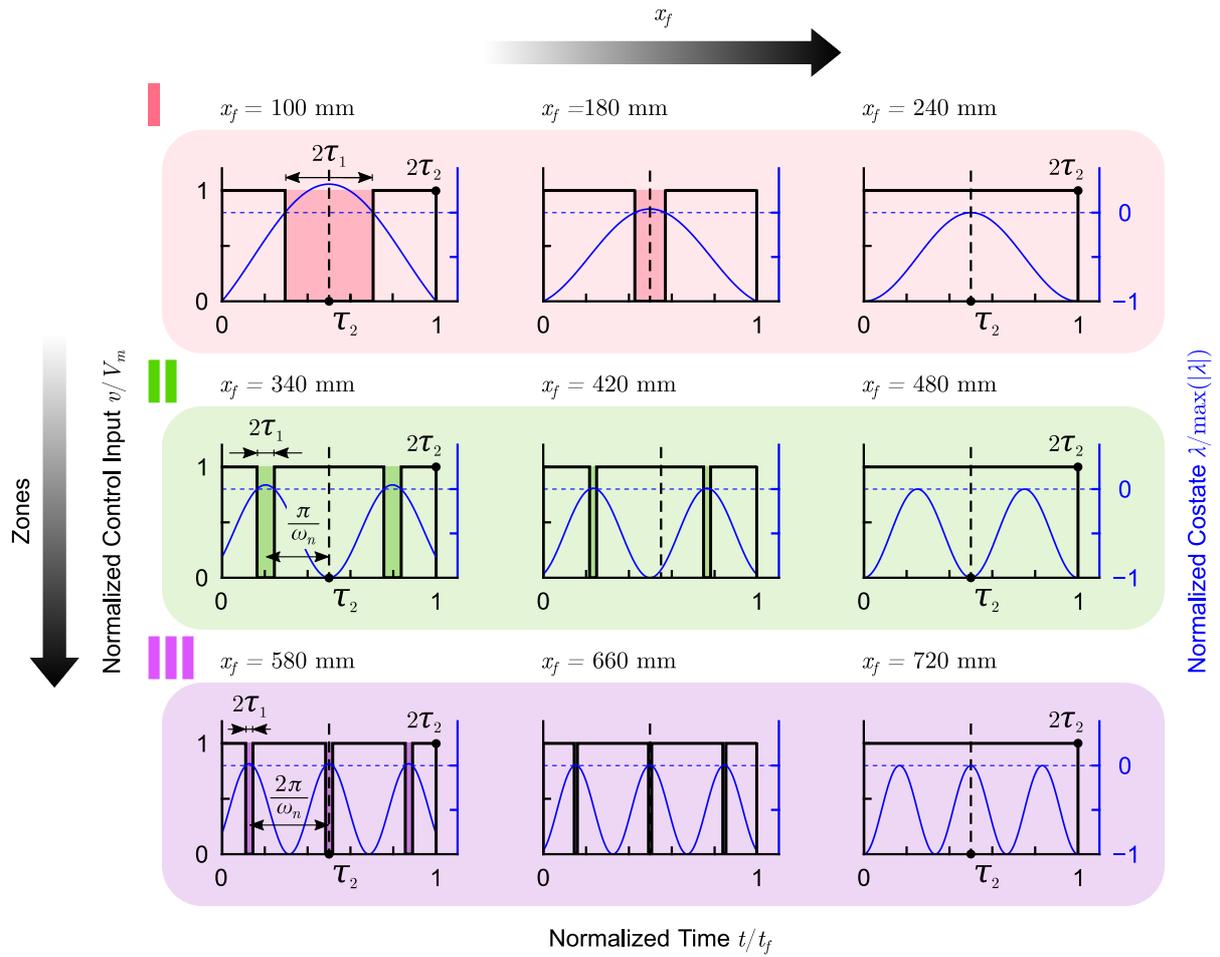}
	\caption{Switching structure and switch time variation as a function of $x_f$ and $\tau_i = T_i/t_f$. Shaded areas illustrate the time span of $T_1$ and the blue curve shows the switching function. Different background colors illustrate the three different picked Zones for (I) Zone 1, (II) Zone 2, (III) Zone 3.}
	\label{fig:03}
\end{figure*}
Since all the off zones of the bang-off-bang profiles are the same, we parameterize the width of each zero zone as $2T_1$ and only two parameters are needed to determine the velocity limited time-optimal control profile. Fig.~\ref{fig:03} illustrates the variation of the switch time for each zone as the terminal displacement increases. The change in structure of the time-optimal control profile with increasing terminal displacements is illustrated for three consecutive zones. Zone 1 is characterized by two switches which collapse for a specific terminal displacement. Any increase in terminal displacement results in the introduction of two additional switches in the optimal control profile. Fig.~\ref{fig:03} also plots the switching function which illustrates that as the optimal control profile transitions from one zone to the next. The switching function includes a phase shift of $\pi$ radians, which results in new switches being birthed at the initial and terminal times of the maneuver. For higher zones, switches can be introduced at the mid-maneuver time and $2\pi/\omega_n$ distance from each other.

To determine the values for $T_1$, which is half the width of each zero velocity zone, and $T_2$, which is half of the maneuver time, a parameter optimization problem is solved which minimizes $T_2$ subject to the constraints:
\begin{align}
T_2-nT_1 &= \frac{x_f}{2V_m} \label{eq:dispcon11}\\
(-1)^{n+1} n \sin(\omega_n T_1) - \sin(\omega_n T_2) &= 0 \label{eq:flexcon11}
\end{align}
where $2T_1$ is the width of each of the zero pulse and $2T_2$ is the maneuver time. Eq.~\eqref{eq:dispcon11} is derived from the final displacement constraint and Eq.\eqref{eq:flexcon11} is derived from the constraint that the time-delay filter needs to locate a pair of zeros at the location of the undamped poles of the system. Eqs.~\eqref{eq:dispcon11} and \eqref{eq:flexcon11} can be transformed into polynomial equations which can be solved efficiently to identify the optimal parameters $T_1$ and $T_2$. The Appendix provides the details of the transformation to generate polynomial equations for zones 2 and 3 for illustrative purposes.

It is interesting to note that as $x_f \rightarrow 0$, Eq.~\eqref{eq:dispcon11} requires $T_1\rightarrow T_2$. Eq.~\eqref{eq:flexcon11} where $n=1$ requires
\begin{equation}
\omega_n T_2 = \pi - \omega_n T_1
\end{equation}
which results in the equation:
\begin{equation}
2T_2 = \frac{\pi}{\omega_n}
\end{equation}
which one can note is analogous to the two-impulse input shaper~\cite{Singer.1990} which for an undamped system places the two impulses of equal magnitude half a period of oscillation apart. Note that $2T_2$ corresponds to the maneuver time. 

\begin{figure}[thpb]
	\centering
	\includegraphics[width=0.45\textwidth]{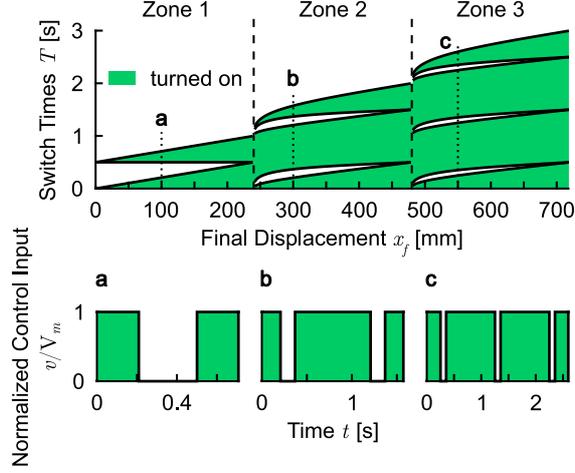}
	\caption{Switch and maneuver time variation of a non-robust time-optimal controller for an undamped 1 mode system. (a), (b) and (c) show the control profiles for $x_f=100$ mm, $x_f=300$ mm and $x_f=550$ mm in  Zone 1, Zone 2 and Zone 3 respectively.}
	\label{fig:04}
\end{figure}

The upper graph of Fig.~\ref{fig:04} illustrates the variation of the switch times and the maneuver time as a function of the terminal displacement $x_f$. The solid colored zone corresponds to the time intervals when the input (velocity command) is at the maximum and the rest of the time intervals are when the input is zero. When $x_f=240$ mm, the two switches collapse resulting in a constant velocity profile. This is followed by the birth of four switches which collapse concurrently for a terminal displacement of $x_f=480$ mm following which a six switch optimal control profile is birthed. The lower half of Fig.~\ref{fig:04} illustrates the optimal solution for three unique displacements in zones 1, 2 and 3, labelled ``a'', ``b'', and ``c'' respectively.
\section{Robust Control}
\label{sec:robcon}
\begin{figure}[thpb]
	\centering
	\includegraphics[width=0.45\textwidth]{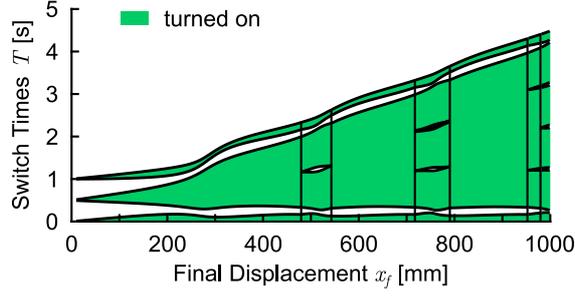}
	\caption{Switch and maneuver time variation of a robust time-optimal controller for an undamped 1 mode system.}
	\label{fig:05}
\end{figure}

\begin{figure}[thpb]
	\centering
	\includegraphics[width=0.45\textwidth]{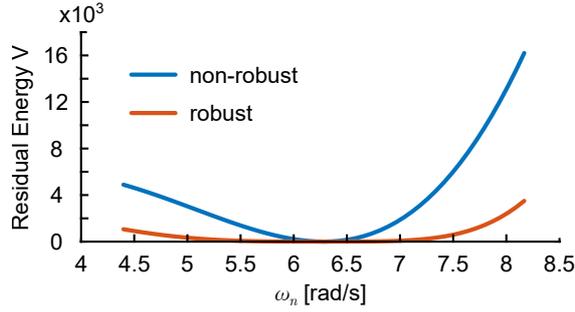}
	\caption{Residual energy at $t_f$ of a non-robust and robust time-optimal controller for a perturbation of $\pm 30$ \% in $\omega_n$ at $x_f=50$ mm.}
	\label{fig:06}
\end{figure}

\begin{figure}[thpb]
	\centering
	\includegraphics[width=0.45\textwidth]{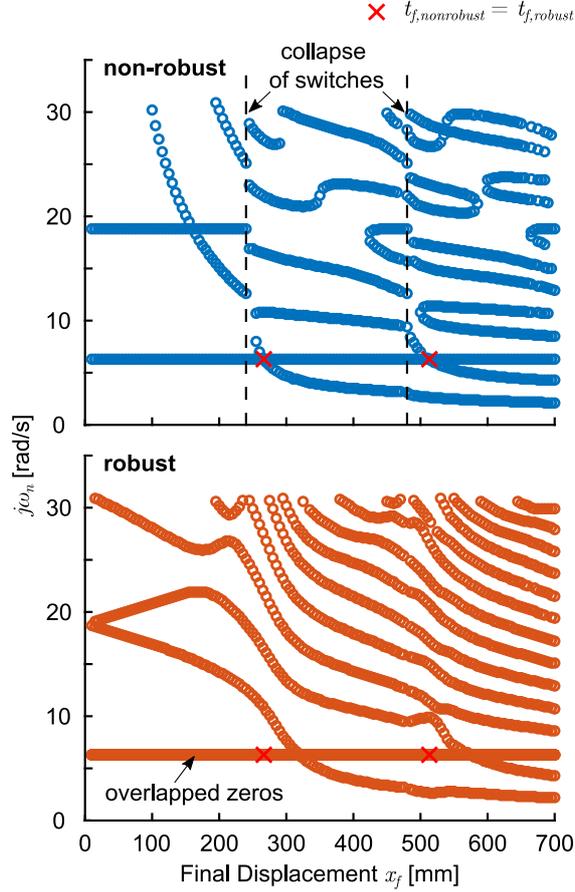}
	\caption{Loci of zeros of the optimal time-delay filter for a non-robust and robust controller as a function of $x_f$ for a 1 mode system.}
	\label{fig:08}
\end{figure}

The challenge of dealing with model parameter uncertainties is ubiquitous and there have been numerous approaches proposed for the design of robust open-loop controllers including enforcing robustness around the nominal model of the system or a minimax problem formulation where the maximum residual energy is minimized over an interval of uncertainty. In this research we determined the sensitivity of the states of the system with respect to uncertainty in the spring stiffness which correspond to uncertainties in the natural frequency and force the state sensitivities with respect to the uncertain frequency to zero at the terminal time. The resulting augmented state space model is:
\begin{align}
\dot{x}_1(t) &= x_2(t)\\
\dot{x}_2(t) &= -\omega_n^2x_1(t) + \omega_n^2x_3(t)\\
\frac{d\dot{x}_1(t)}{d\omega_n} &= \frac{dx_2(t)}{d\omega_n}\\
\frac{d\dot{x}_2(t)}{d\omega_n} &= -2\omega_n x_1(t) -\omega_n^2\frac{dx_1(t)}{d\omega_n}+ 2\omega_n x_3(t)\\
\dot{x}_3(t) &= v(t)\\
0 & \leq v \leq V_m.
\end{align}
and is subject to the initial and final conditions:
\begin{align}
x_1(0)&=x_2(0)=x_3(0)=0\\
\frac{dx_1(0)}{d\omega_n}&=\frac{dx_2(0)}{d\omega_n}=0 \\
x_1(t_f) &= x_3(t_f)= x_f\\
x_2(t_f)&=\frac{dx_1(t_f)}{d\omega_n}=\frac{dx_2(t_f)}{d\omega_n}=0.
\end{align}
The robust velocity constrained time-optimal control problem for the undamped system is solved for various displacements.
The variation in the optimal control profile parameterized by the switch times and maneuver time are illustrated in Fig.~\ref{fig:05} as a function of the terminal displacement. Unlike the variation of the switch times as a function of terminal displacement and the concurrent collapse of the switches, this phenomena is not observed in the spectrum of switch times in Fig.~\ref{fig:05}. It nevertheless should be noted that the optimal control profile births and collapses switches as a function of terminal displacements, resulting in an optimal control profile where the number of switches is a function of the terminal displacement.
Fig.~\ref{fig:06} illustrates the variation of the residual energy at the terminal time for the non-robust and robust time-optimal controllers over a range of uncertain natural frequencies for the undamped system. It is clear that the red line, which represents the variation of residual energy of the robust control, outperforms the non-robust design illustrated by the blue line. These graphs are generated for a terminal displacement of $x_f=50$ mm and a natural frequency of $\omega_n = 2\pi$.
The residual energy is given by:
\begin{align}
V(t) = \frac{1}{2}\dot{x}(t)^2 + \frac{1}{2}\omega_n^2\left(x(t)-x_f\right)^2. \label{eq:residual_energy}
\end{align}
The sensitivity of the residual energy with respect to the natural frequency is:
\begin{align}
\frac{dV(t)}{d\omega_n} &= \dot{x}(t)\frac{d \dot{x}(t)}{d\omega_n} + \omega_n\left(x(t)-x_f\right)^2 + \omega_n^2\left(x(t)-x_f\right) \frac{dx(t)}{d\omega_n}. \label{eq:residual_energy_sen}
\end{align}
Since the time-optimal control profile forces the terminal states to be $x(t_f)=x_f$ and $\dot{x}(t_f)=0$, $V(t_f)$ and $\frac{dV(t_f)}{d\omega_n}=0$, irrespective of the magnitude of the sensitivity states, $\frac{dV(t)}{d\omega_n}=0$ at the terminal time. The second derivative of the residual energy with respect to the natural frequency is:
\begin{align}
\frac{d^2V(t)}{d\omega_n^2} &= \left(\frac{d\dot{x}(t)}{d\omega_n}\right)^2 + \dot{x}\frac{d^2\dot{x}(t)}{d\omega_n^2} + \left(x(t)-x_f\right)^2 + 4\omega_n\left(x(t)-x_f\right)\frac{dx(t)}{d\omega_n} + \omega_n^2\left(\frac{dx(t)}{d\omega_n}\right)^2 ...\nonumber\\ 
&... + \omega_n^2\left(x(t)-x_f\right)\frac{d^2x(t)}{d\omega_n^2}
\end{align}
which is the curvature since the first derivative is zero at the nominal frequency $\omega_n$. The curvature is defined as the reciprocal of the radius of a circle which best approximates the curve $V(\omega_n)$. The curvature can be used as a measure of the robustness of the control profile in the proximity of the nominal model with a smaller curvature (i.e. an osculating circle with large radius) indicating a smaller residual energy variation or greater robustness to uncertainties in the natural frequency.
\begin{figure}[thpb]
	\centering
	\includegraphics[width=0.45\textwidth]{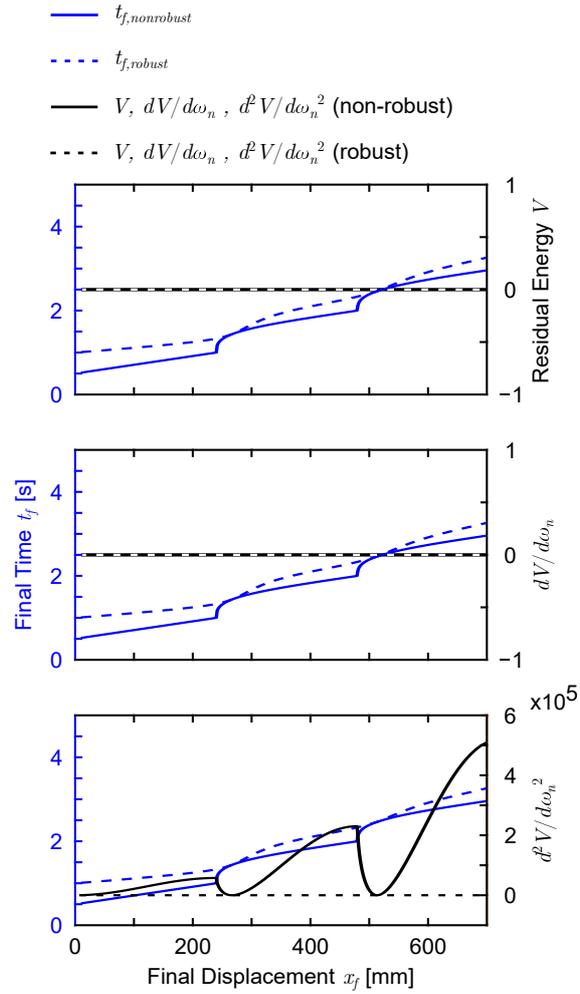}
	\caption{Variation of terminal time, residual energy, and its derivatives as a function of $x_f$ for a non-robust and robust time-optimal controller to illustrate the collapses of the switch times for an undamped 1 mode system.}
	\label{fig:07}
\end{figure}
The solid and dashed lines in Fig.~\ref{fig:07} correspond to the respective non-robust and robust solutions, the blue curves represent the variation in the maneuver time as a function of terminal displacement, and the black lines represent the variation of the residual energy and its derivatives with terminal displacement.
The third panel of Fig.~\ref{fig:07} shows that there is a profound variation in the curvature of residual energy function for the non-robust solution as a function of terminal displacement in comparison to that of the robust control which is zero for all displacements. It is also intriguing that, for specific terminal displacements, the robust and non-robust solutions are identical, i.e., the maneuver times are the same and the curvatures are both zero.
To comprehend this unique result the loci of the zeros of the time-delay filter which generates the bang-off-bang control profiles are generated and are illustrated in Fig.~\ref{fig:08}.
\begin{figure}[thpb]
	\centering
	\includegraphics[width=0.45\textwidth]{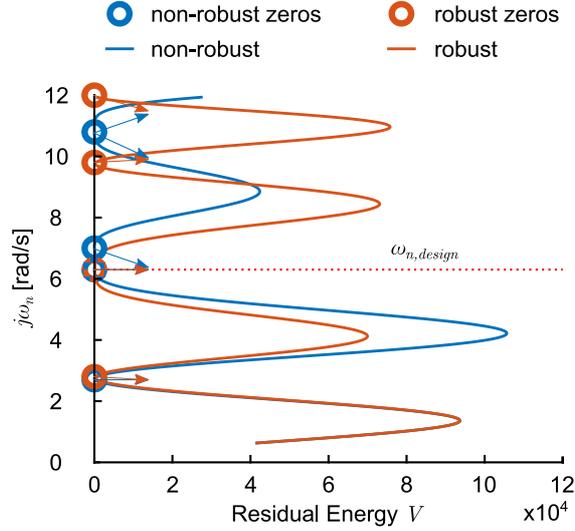}
	\caption{Residual energy versus uncertain model frequency for a non-robust and robust controller at $x_f=500$ mm for a 1 mode system. The arrows indicate the slopes of zero-loci for $x_f > 500$ mm. $\omega_{n,design}$ shows the frequency the time-delay filters were designed for.}
	\label{fig:09}
\end{figure}

All the graphs are generated for a nominal frequency of $\omega_n = 2\pi$. 
As illustrated by the upper graph of Fig.~\ref{fig:08}, the non-robust design requires that the time-delay filter place zeros at the location of the nominal undamped poles of the plant.  Meanwhile, the lower graph of Fig.~\ref{fig:08} shows that the robust design mandates that a pairs of zeros be located at the nominal poles of the plant. 
The horizontal loci of zeros for both the robust and non-robust designs illustrates that the time-delay filter places single and double sets of zeros at the nominal location of the poles for the non-robust and robust designs respectively. It can also be noted that the zero loci for the non-robust design intersects the horizontal loci for a terminal displacement of $x_f=266.5$ mm (shown by the red cross), resulting in a pair of coincident zeros which is what the robust-controller is mandated to do. This is the reason why, for specific terminal displacements, the robust and non-robust designs results in identical solutions. It can also be seen that the same phenomena occurs for a terminal displacement of $x_f=513.3$ mm. 

Fig.~\ref{fig:09} illustrates the location of the zeros of the time-delay filter for the non-robust and robust designs for the terminal displacement $x_f=500$ mm. The blue circles correspond to the non-robust design and the red to the robust design. It should be noted that the robust design includes two pairs of zeros at $\pm 2\pi j$ and the non-robust design includes one pair of zeros. The arrows associated with the blue and red circles are the slopes of the loci of the zeros as a function of terminal displacement $x_f$. It is  evident that a blue zero is transitioning from above $2\pi j$ with a negative slope and is shown to coincide with the nominal poles of the system for $x_f=513.3$ mm, resulting in identical robust and non-robust designs. 

The red and blue lines in Fig.~\ref{fig:09} illustrate the variation in residual energy as the location of the nominal poles of the system are varied. The non-robust design outperforms the robust design for perturbation of the uncertain frequency above $\omega_n=2\pi$, while the robust design outperforms the non-robust design for perturbations below the nominal frequency. This asymmetry is due to the fact that the non-robust design has a zero located immediately above the nominal frequency and ends up acting like the zero locations of a minimax design~\cite{Singh.2002}. It can also be seen that, for large perturbations of the uncertain frequency, the residual energy goes to zero at locations where the time-delay filter has a zero.

\section{Generalization}
\label{sec:robdesign}
Section~\ref{sec:prob_st} dealt with an undamped system with one mode which permitted a reduced order parameterization of the optimal control profile by exploiting the symmetrical nature of the control about the mid-maneuver time. This symmetry is attributed to the fact that the oscillatory motion excited by the input does not damp out and the symmetric input can, by virtue of linearity of the system, generate an out of phase motion of the undamped modes which cancels the existing oscillations. For a system with damping, the symmetric nature of the control profile is lost since the amplitude of the oscillatory mode decays over time. Consequently, the optimal control profiles need to explicitly parameterize every switch in addition to the maneuver time. Furthermore, there may be multiple modes which contribute to the output of interest and those modes need to be quiescent at the end of the maneuver as well. For a system with multiple modes, characterized by damped or undamped modes, the bang-off-bang control profile is parameterized as:
\begin{align}
v(s) = \frac{V_m}{s}\underbrace{\left[ 1 + \sum_{i=1}^{N+1} (-1)^i e^{-s T_i} \right]}_{=G_c(s)}
\end{align}
where $N$ is the number of switches which is an even number and $T_{N+1}$ is the maneuver time. 

The constraints to identify the optimal values of $T_i$ are derived by requiring the zeros of the time-delay filter to cancel the pole at the origin and the underdamped poles of the system. This results in the constraint:
\begin{align}
G_c(s=0) = 1 + \sum_{i=1}^{N+1} (-1)^i  = 0
\end{align}
which is automatically satisfied by the parameterization of the optimal control profile. To cancel the complex conjugate poles at $s=-\sigma_k \pm j\:\omega_{d,k} = -\zeta_k\omega_{n,k}\pm j\:\omega_{n,k}\sqrt{1-\zeta_k^2}$, where $k=1,2,\hdots,m$ are the $m$ modes whose poles need to be canceled by the zeros of the time-delay filter. This results in the constraints:
\begin{align}
1 + &\sum_{i=1}^{N+1} (-1)^i e^{\sigma_k T_i}\cos(\omega_{d,k} T_i) = 0\\
&\sum_{i=1}^{N+1} (-1)^i e^{\sigma_k T_i}\sin(\omega_{d,k} T_i) = 0,\: k=1,2,\hdots,m.\label{eq:sin_underdamped_unsimplified}
\end{align}
To satisfy the terminal rigid body displacement, we require:
\begin{align}
x(T_{N+1}) &= x_f = \int^{t_f}_0 V_m dt \\ 
&= V_m \left(T_{N+1} + \sum_{i=1}^{N} (-1)^i  (T_{N+1}-T_i) \right)\\
\rightarrow x_f &= V_m \left( T_{N+1}-\sum_{i=1}^{N} (-1)^i T_i \right).
\label{eq:termdisp_5_switches}
\end{align}
The nonlinear optimization problem that requires solving is:
\begin{subequations}
	\begin{align}
	& \mbox{min} ~~J = T_{N+1} \\
	\mbox{subject to} & \notag \\ & V_m \left( T_{N+1}-\sum_{i=1}^{N} (-1)^i T_i \right) = x_f \\
	& 1 + \sum_{i=1}^{N+1} (-1)^i e^{\sigma_k T_i}\cos(\omega_{d,k} T_i) = 0 \\
	& \sum_{i=1}^{N+1} (-1)^i e^{\sigma_k T_i}\sin(\omega_{d,k} T_i) = 0,\: k=1,2,\hdots,m \\
	& 0 \le (T_{i+1}-T_i), \mbox{  for  } i=0,1,2,\hdots,N
	\end{align}
\end{subequations}
where $T_0=0$.
\begin{figure}[ht]
	\centering
	\includegraphics[width=0.45\textwidth]{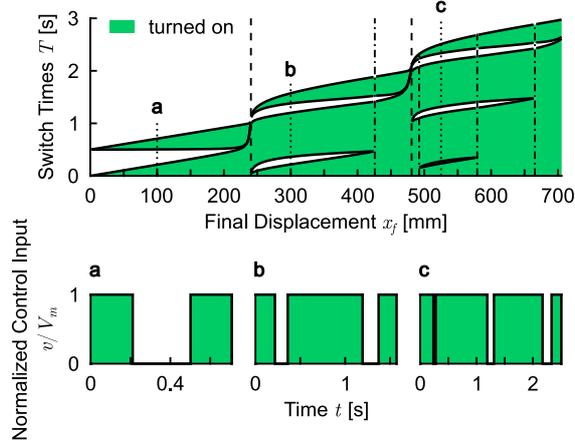}
	\caption{Switch and maneuver time variation of a non-robust time-optimal controller for an underdamped 1 mode system.}
	\label{fig:10}
\end{figure}
As in the case of the undamped system, the number of switches necessary to parameterize the optimal control profile changes with the terminal displacement. Fig.~\ref{fig:10} illustrates the variation of the switch times as a function of the final displacement and the control input over the maneuver time for a single mode system with an underdamped system model ($\zeta=0.01$). For terminal displacements of $0-700$ mm, three zones have been identified, separated by the vertical dashed lines. Three representative optimal control profiles are presented corresponding to three displacements highlighted by the dotted lines and labelled $a$, $b$, and $c$. It is interesting to note that in zone 3, which includes the dotted line $c$, the number of switches required for the optimal control profile starts with four switches, transitions to six switches, and returns to a four switch optimal control profile before finally transitioning to a two switch profile.

\section{Switching Profile Transition}
\label{sec:sw_tran}
It is clear from Figs.~\ref{fig:04} and \ref{fig:10} that the structure of bang-off-bang control profile changes as a function of the terminal displacement. In Fig.~\ref{fig:04}, for small displacements ($x_f \le 240$ mm), the optimal control profile is first characterized by two switches and the switches, then, after collapsing, result in a pulse control profile which births a four switch control profile which subsequently transitions to a six switch control profile. The change in structure of the optimal control profile for underdamped systems is more involved as all switches do not collapse for the same value of the terminal displacement. To exactly determine the terminal displacement which corresponds to the birth or collapse of two or more switches, the constraint is that the switching function and its time derivative are simultaneously zero at some time instant. For the underdamped system the switching function $\lambda_3(t)$ can be represented as:
\begin{eqnarray}
    \lambda_3(t_{cr},x_f) = \dot{\lambda}_3(t_{cr},x_f) = 0
\end{eqnarray}
where $t_{cr}$ is the switch time where two switches collapse. $t_{cr}$ and $x_f$ can be determined by solving the two nonlinear simultaneous equations in two unknowns while satisfying all the necessary conditions for optimality.

\section{Experimental Results}
\label{sec:experiment}
\begin{figure}[thpb]
\centering
\includegraphics[width=0.5\textwidth]{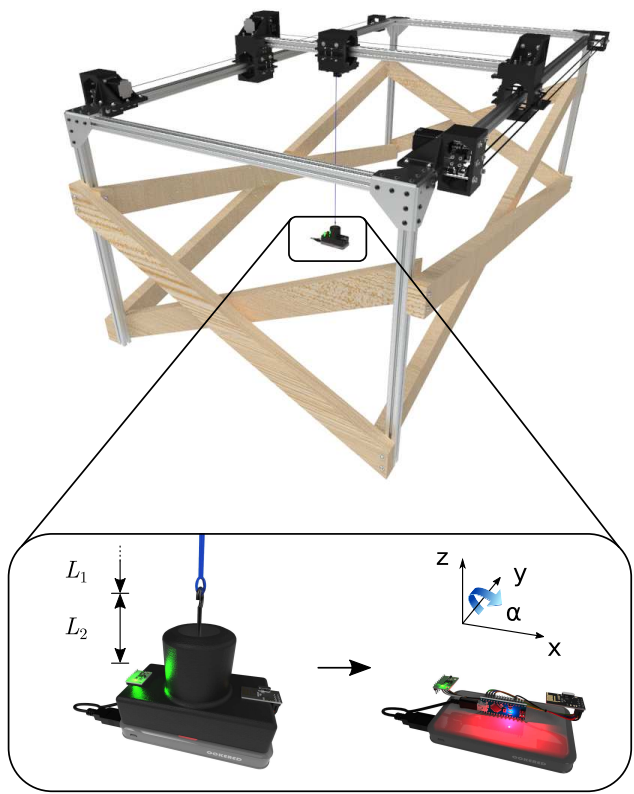}
	\caption{Experimental setup of a gantry crane with an inset of the sensor part, which consists of a powerbank, Arduino Nano, 3-axis gyroscope (MPU-6050), and an nRF24L01 transceiver. Rotation around the y-axis is introduced as the swinging angle $\alpha$. $L_1$ is denoted as the rope length and $L_2=7.86$ cm}
 \label{fig:11}
\end{figure}

A scaled model of a gantry crane was fabricated to test and validate the time-optimal control profiles presented in this paper. The workspace of the gantry crane includes a cuboid of dimensions $8$' $\times$ $4$' $\times$ $3$' as shown in Fig.~\ref{fig:11}. Its main element is a trolley which slides on a rail and is able to move a payload over a $2$-dimensional space (the winch motion of the crane is not active). The crane is equipped with $3$ stepper motors of type NEMA 23 with a driver setting of $400$ steps per revolution, where two of them are used for the y-axis and one for the x-axis. For this experiment just the motor of the x-axis is used. The drivers of the stepper motors are connected to an Arduino MEGA $2560$ which receives the switch times and the desired position of the trolley as inputs. The maximum velocity of the trolley is set to $240$ mm/s.
The inset of Fig.~\ref{fig:11} includes two images. The one on the left illustrates the 3D printed chassis which houses a cylindrical steel mass of $500$g. A rope and a hook enable a connection between the chassis and the trolley making it a double pendulum system. The right panel illustrates the sensor integrated into the payload which includes two Arduino Nanos, a 3-axis gyroscope (MPU-6050), and two nRF24L01 single chip radio transceivers. One Arduino Nano is housed within the chassis to process the data provided by the gyroscope and to transmit the data to the other Arduino Nano, which is connected to a receiver. To permit a repeatable evaluation of the robustness of the controller, a cable deployment setup was designed which permits changing the length $L_1$ of the pendulum about a nominal length. The change in cable length changes the modal frequency of the system and was used to test the robustness of the controller to uncertainties in modal parameters of the gantry crane system. 

A series of experiments were conducted to illustrate some of the novel observations of the analytical study. The first of which included the collapse of the switches of the velocity constrained time-optimal control profile for a system with a single undamped mode. The simplest model of the gantry crane includes representing the suspended mass as a single undamped pendulum and a rectangular command, i.e., a bang control profile should cancel the undamped mode for a specific displacement. The natural frequency and damping ratio of the first mode of the pendulum with the nominal length were experimentally determined to be $\omega_n = 0.6832$ Hz and $\zeta_1= 0.001517$ which was approximated to be zero in the development of the controller. 

\begin{figure}[thpb]
\centering
	\includegraphics[width=0.5\textwidth]{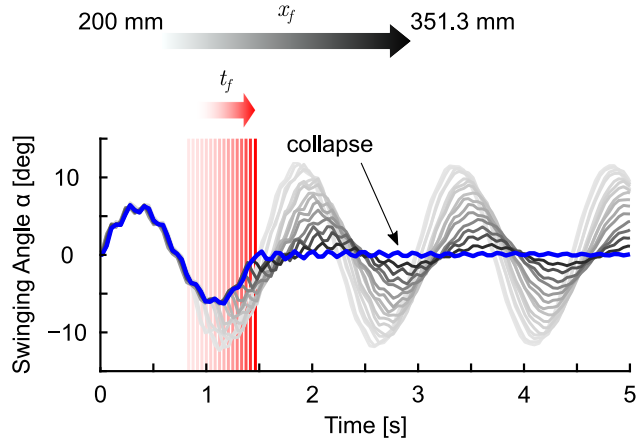}
	\caption{Residual vibration variation when changing the final displacement from $x_f=200$ mm to $x_f=351.3$ mm for a pulse control profile. The blue curve illustrates the residual vibration at $x_f=351.3$ mm which validates the cancellation of the first mode of vibration.}
 \label{fig:12}
\end{figure}

\begin{figure}[thpb]
\centering
	\includegraphics[width=0.45\textwidth]{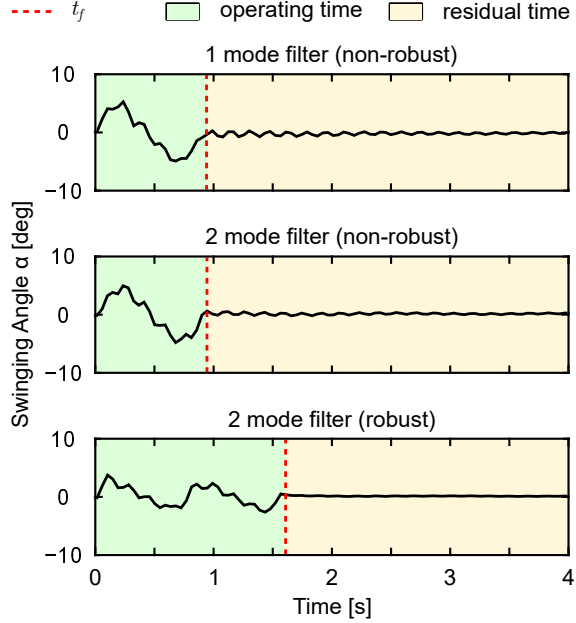}
	\caption{Swinging angle $\alpha$ (around y-axis) for a 1 mode time-delay filter (non-robust), 2 mode time-delay filter (non-robust), and 2 mode time-delay filter (robust) when $x_f=100$ mm. Shaded regions help to distinguish between the operating and residual (off) time of the controller. The final time $t_f$ for each scenario is highlighted by a red dashed line.}
 \label{fig:13}
\end{figure}

\begin{figure}[thpb]
\centering
	\includegraphics[width=0.5\textwidth]{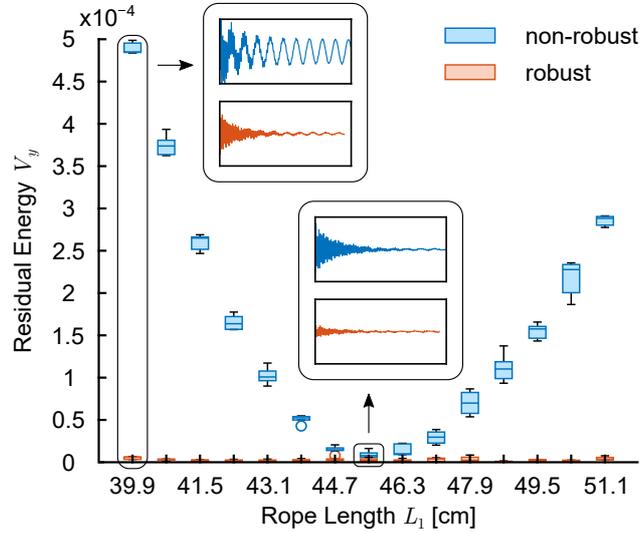}
	\caption{Residual energy variation at $t_f$ of time-optimal controllers for $x_f=100$ mm for different rope lengths $L_1$ varying from $39.9$ cm to $51.1$ cm in 15 equally spaced intervals. The time-optimal controllers are designed for $L_1=45.5$ cm. Blue and red shaded illustrate the residual energy statistics for the 2 mode non-robust and robust controller respectively. Furthermore, the time history of the swinging angle $\alpha$ for a non-robust vs robust scenario is illustrated by insets. The overall box and whisker chart includes 150 experiments.}
 \label{fig:14}
\end{figure}

\begin{figure}[thpb]
\centering
\includegraphics[width=0.5\textwidth]{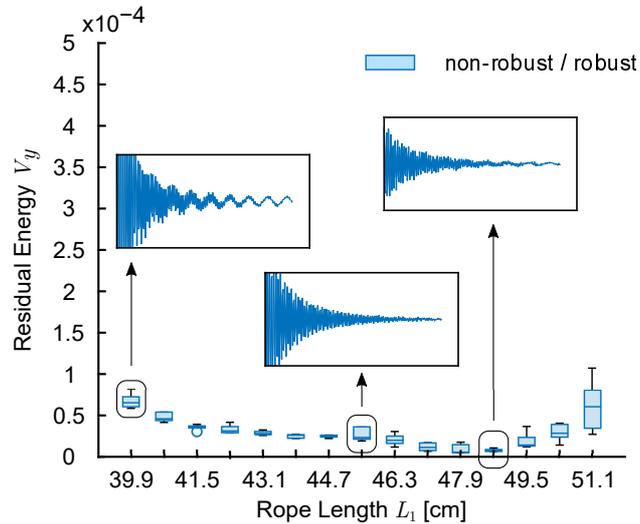}
	\caption{Residual energy at $t_f$ of a time-optimal controller at $x_f=390.1$ mm for different rope length $L_1$ varying from $39.9$ cm to $51.1$ cm in 15 equally spaced intervals. The controller is designed for a system with $L_1=45.5$ cm where the non-robust and robust solution collapse. The change of swinging angle $\alpha$ for the controller is illustrated by insets. The overall box and whisker chart includes 75 experiments.}
 \label{fig:15}
\end{figure}

Fig.~\ref{fig:12} illustrates experimental results for terminal displacements resident in zone 1, which corresponds to the optimal control profile being characterized by a two switch bang-off-bang profile. A pulse with increasing width is progressively applied to illustrate the fact that for specific terminal displacements, a bang profile (pulse) results in zero terminal vibration. Starting from a terminal displacement of $x_f=200$ mm and progressively increasing it until $x_f=351.3$ mm, the maneuver time increases, but at the same time, the maximum angular displacement of the pendulum $\alpha$ is decreasing until it is not present at the end of the maneuver for a displacement of $351.3$ mm. It can be noted that a small high frequency oscillation is evident which corresponds to the second mode of the double pendulum system which is not considered in the controller design.
For the next set of experiments, the double pendulum model of the crane is assumed in the identification of the parameters of the two modes of oscillation. The second mode's parameters are identified to be  $\omega_{n,2} = 6.159$ Hz and $\zeta_2 = 0.026065$ where the damping of the second mode is an order of magnitude greater than that of the first. For the two mode system, three controllers were tested. The first controller was designed to cancel the first mode only. The second controller was designed to cancel both of the modes at the end of the maneuver and the third controller was designed to be robust to uncertainties in both the modes' natural frequencies. Fig.~\ref{fig:13} illustrates the time variation of the angular displacement of the pendular payload where the green region corresponds to when the control input is active and the yellow region is the post-actuation domain. The first panel of Fig.~\ref{fig:13} illustrates the impact of cancelling the first mode, the second panel corresponds to the controller cancelling both modes, and the third panels display the elimination of residual vibration when the robust time-optimal control profile is used to drive the system. These results were generated for a terminal displacement of $100$ mm and the pendulum length $L_1$ which corresponds to the nominal model which was used to identify the modal parameters.
To illustrate the variation in residual energy as the pendular length is varied, five experiments were conducted for each pendular length to account for uncertainties in initial conditions. With a total of seven perturbed lengths of the pendulum on either side of the nominal length, a total of 150 experiments were conducted. Box and whisker charts are used to illustrate uncertainties in the residual energy for each of the perturbed models. Fig.~\ref{fig:14} illustrates the profound improvement in the use of the robust controller for which the residual vibration for the various pendulum lengths appear negligible compared to the residual energy resulting from the use of the non-robust control profiles.
The final set of experiments illustrates a rather counter intuitive result which claims that the non-robust design and robust designs are coincident since the non-robust design for specific terminal displacements of the single-mode model places multiple zeros of the time-delay filter at the nominal location of the poles of the system.
Fig.~\ref{fig:15} illustrates the residual vibration box and whisker charts for a terminal displacement of $x_f = 390.1$ mm, where it is evident that the residual energy curve is relatively flat. It should be pointed out that in this design, only the first mode of vibration was considered in the design, since observation of the coincidence of the robust and non-robust design has been observed for system with single undamped modes.
\section{Conclusions}
This paper presents an optimal control based development of a velocity limited minimum time control of a gantry crane system which is characterized by two modes of vibratory motion. The variation in the structure of the optimal control profile is presented for a single mode system where the vibratory modes are undamped or underdamped. It is noted that, as the final displacement increases, there is an increase in the number of switches in the optimal control profile with periodic terminal displacements requiring a pulse control profile with no switches. The optimal control framework is extended to account for multiple vibratory modes such as when the crane is modeled as a double pendulum. The state sensitivities are used to determine controllers which are robust to uncertainties in model parameters. Experimental results validate the counter intuitive observation that for specific terminal displacement the robust and non-robust optimal control profiles are coincident. The experimental results also clearly demonstrate the profound reduction in residual vibrations of the double pendulum system when the length of the pendulum is changed to serve as a proxy for uncertainties in natural frequencies of the model.

\section*{Acknowledgment}
The authors acknowledge the support of this work by the US National Science Foundation through CMMI Award number 2021710. The authors would like to thank Dr. Claude F. Leibovici for his help in transforming Eq.~\eqref{eq:dispcon11} and \eqref{eq:flexcon11} to a polynomial equation. 

\vspace{-0.1in}
\bibliographystyle{elsarticle-num}
\bibliography{ref} 
\noindent\rule[0.5ex]{\linewidth}{1pt}

\newpage
\onecolumn

\appendix
\section*{Appendix}
\renewcommand{\theequation}{A-\arabic{equation}}     
\setcounter{equation}{0}
The time-optimal control profile for zone 1 illustrated in Fig.~\ref{fig:04} parameterized with two variables $T_1$ and $T_2$ has a closed form solution. For higher order zones, the transcendental constraint equations cannot be solved in closed form, but can be converted to a polynomial equation as illustrated for zones 2 and 3.
\subsection*{Zone 2}
The parameterization of the time-optimal control for Zone 2 leads to the constraints:
\begin{align}
    2\sin(\omega_n T_1) + \sin(\omega_n T_2) = 0 \label{eq:3_switches_eq1}\\
    2T_2 - 4T_1 = \frac{x_f}{V_m}. \label{eq:3_switches_eq2}
\end{align}
Let $x_f/V_m=2a\omega_n$, which gives $2T_2-4T_1=2a\omega_n$ or $T_2=2T_1+a\omega_n$. Eq.~\eqref{eq:3_switches_eq1} becomes $2\sin(\omega_n T_1) +  \sin(2\omega_nT_1+a\omega_n^2)=0$. This can be further simplified to:
\begin{align}
    2\sin\left(\omega_n T_1\right) +  \sin\left(2\omega_nT_1\right)\cos\left(a \omega_n^2\right) + \sin\left(a\omega_n^2\right)\cos\left(2\omega_n T_1\right)=0.
\end{align}
Let $\alpha = \sin\left(a\omega_n^2\right)$, $\beta=\cos\left(a\omega_n^2\right)$ and $t=\omega_nT_1$:
\begin{align}
    2\sin(t) + \beta\sin(2t) + \alpha\cos(2t) = 0.
\end{align}
Let $t=\cos^{-1}(z)$, resulting in the simplified equation:
\begin{align}
    2\sin(\cos^{-1}(z)) + \beta \sin(2cos^{-1}(z)) + \alpha \cos(2cos^{-1}(z)) &= 0\\
    2\sqrt{1-z^2} + \beta 2z\sqrt{1-z^2} + \alpha \left(2z^2-1\right) &= 0\\
    2\sqrt{1-z^2}\left(\beta z + 1\right) + \alpha \left(2z^2-1\right) &= 0\\
    \alpha\left(2z^2-1\right) &= -2\sqrt{1-z^2}\left(\beta z + 1\right)\\
    \alpha^2\left(4z^4-4z^2+1\right) &=4\left(1-z^2\right)\left(\beta^2z^2+2\beta z + 1\right)\\
    \left(4\alpha^2+4\beta^2\right)z^4 + 8\beta z^3 + \left(-4\alpha^2-4\beta^2+4\right)z^2 - 8\beta z + \alpha^2 - 4 &= 0.
\end{align}
Exploiting the knowledge that $\alpha^2+\beta^2=1$, we have:
\begin{align}
    4z^4 + 8\beta z^3 - 8\beta z + \alpha - 4 = 0. \label{eq:3_switches_polynomial}
\end{align}
From Eq.~\eqref{eq:3_switches_polynomial} the parameter $T_1$ can be calculated by using $t=cos^{-1}(z)$ and $T_1=t/\omega_n$. Assuming that $\omega_n$ is given and the user can choose any $x_f$ which lies in the bounds of the zone, the quartic equation provides a solution for the switch time $T_1$. From there, the mid-maneuver time $T_2$ can easily be calculated by Eq.~\eqref{eq:3_switches_eq2}. For illustrative purposes, assume $\omega_n=2\pi$, $x_f=400$ mm and $V_m=240$ mm/s. Since the discriminant of the quartic equation is negative, we have two real and two complex conjugate roots. We disregard the complex roots, obtain $T_{1,1} = 0.0409$ s and $T_{1,2} = 0.4247$ s, and from there $T_{2,1} =0.9151$ s and $T_{2,2} = 1.6827$ s follows. The time-optimal problem requires the shorter time which is why $T_{2,2}$ is discarded.
\subsection*{Zone 3}
The switch parameterization for Zone 3 can be written as:
\begin{align}
    3\sin(\omega_n T_1) - \sin(\omega_n T_2) = 0 \label{eq:5_switches_eq1}\\
    2T_2 - 6T_1 = \frac{x_f}{V_m}. \label{eq:5_switches_eq2}
\end{align}
Let $x_f/V_m=2a\omega_n$, which gives $2T_2-6T_1=2a\omega_n$ or $T_2=3T_1+a\omega_n$. Eq.~\eqref{eq:5_switches_eq2} becomes $-3\sin\left(\omega_n T_1\right) +  \sin\left(2\omega_n T_1+a\omega_n^2+\omega_nT_1\right)=0$. This can be further simplified to:
\begin{align}
    -3\sin\left(\omega_n T_1\right) +  \sin\left(2\omega_nT_1\right)\cos\left(a\omega_n^2+\omega_nT_1\right) + \sin\left(a\omega_n^2 + \omega_nT_1\right)\cos\left(2\omega_n T_1\right)=0. 
    \label{eq:AppZ3_1}
\end{align}
Let $\alpha = \sin\left(a\omega_n^2\right)$, $\beta=\cos\left(a\omega_n^2\right)$ and $t=\omega_nT_1$, Eq.~\eqref{eq:AppZ3_1} can be rewritten as:
\begin{align}
    -3\sin(t) + \sin(2t)\left(\beta\cos(t)-\alpha\sin(t)\right) + \cos(2t)\left(\alpha\cos(t)+\beta\sin(t)\right) = 0.
\end{align}
Let $t=\cos^{-1}(z)$:
\begin{align}
    -3\sin(\cos^{-1}(z)) + \sin\left(2\cos^{-1}(z)\right)\left(\beta\cos(\cos^{-1}(z)) - \alpha\sin(\cos^{-1}(z))\right)...\nonumber\\
    ...+ \cos(2\cos^{-1}(z))\left(\alpha\cos(\cos^{-1}(z)) + \beta\sin(\cos^{-1}(z))\right) &= 0\\
    -3\sqrt{1-z^2} + 2z\sqrt{1-z^2}\left(\beta z - \alpha\sqrt{1-z^2}\right) + \left(2z^2-1\right)\left(\alpha z + \beta \sqrt{1-z^2}\right) &= 0\\
    -3\sqrt{1-z^2} + 2\beta z^2 \sqrt{1-z^2} - 2\alpha z(1-z^2) + (2z^2-1)\alpha z + \left(2z^2-1\right)\beta\sqrt{1-z^2} &= 0\\
    \sqrt{1-z^2}\left(4\beta z^2-\beta-3\right) &= -4\alpha z^3 + 3\alpha z\\
    \left(1-z^2\right)\left(16\beta^2 z^4 - 8\beta(\beta+3)z^2 + \beta^2 + 6\beta + 9\right) &= 16\alpha^2 z^6 - 24\alpha^2z^4 + 9\alpha^2z^2.
    \end{align}
With the knowledge that $\alpha^2+\beta^2=1$, we have:
\begin{align}
-16z^6 + (24 + 24\beta)z^4 + (-30\beta - 18)z^2 + \beta^2 + 6\beta + 9 = 0 \label{eq:5_switches_polynomial}
\end{align}
which can be reduced to a cubic equation by introducing $z^2 = w$, resulting in the equation:
\begin{align}
-16w^3 + (24 + 24\beta)w^2 + (-30\beta - 18)w + \beta^2 + 6\beta + 9 = 0. \label{eq:5_switches_polynomial_simplified}
\end{align}
If we pick $\omega_n=2\pi$, $x_f=600$ mm  and $V_m=240$ mm/s, the only real solution which would satisfy Eq.~\eqref{eq:5_switches_eq1} is $T_1 = 0.0395$ s and therefore $T_2 = 1.3684$ s. Note, $t=\cos^{-1}(z)$ supports two real solutions and 4 complex conjugate solutions. From there, $T_1$ can be calculated but in this case just one of the real solutions satisfies Eq.~\eqref{eq:5_switches_eq1}. 
\end{document}